\documentclass{aastex631}

\usepackage{amsmath}
\usepackage{amssymb}
\usepackage{etoolbox}
\usepackage{rotating}
\begin{document}

\title{Searching for TeV Dark Matter in Irregular dwarf galaxies with HAWC Observatory}

\correspondingauthor{S.~Hern\'andez-Cadena}
\email{skerzot@ciencias.unam.mx}

\correspondingauthor{V.~Gammaldi}
\email{viviana.gammaldi@uam.es}

\correspondingauthor{J.~Serna-Franco}
\email{j\_serna@ciencias.unam.mx}

\author{R.~Alfaro}
\affiliation{Instituto de F\'{i}sica, Universidad Nacional Autónoma de México, Ciudad de Mexico, Mexico}

\author{C.~Alvarez}
\affiliation{Universidad Autónoma de Chiapas, Tuxtla Gutiérrez, Chiapas, México}

\author{J.C.~Arteaga-Velázquez}
\affiliation{Universidad Michoacana de San Nicolás de Hidalgo, Morelia, Mexico}

\author[0000-0002-4020-4142]{D.~Avila Rojas}
\affiliation{Instituto de F\'{i}sica, Universidad Nacional Autónoma de México, Ciudad de Mexico, Mexico}

\author[0000-0002-2084-5049]{H.A.~Ayala Solares}
\affiliation{Department of Physics, Pennsylvania State University, University Park, PA, USA}

\author{R.~Babu}
\affiliation{Department of Physics, Michigan Technological University, Houghton, MI, USA}

\author[0000-0003-3207-105X]{E.~Belmont-Moreno}
\affiliation{Instituto de F\'{i}sica, Universidad Nacional Autónoma de México, Ciudad de Mexico, Mexico}

\author[0000-0002-4042-3855]{K.S.~Caballero-Mora}
\affiliation{Universidad Autónoma de Chiapas, Tuxtla Gutiérrez, Chiapas, México}

\author[0000-0003-2158-2292]{T.~Capistrán}
\affiliation{Instituto de Astronom\'{i}a, Universidad Nacional Autónoma de México, Ciudad de Mexico, Mexico}

\author[0000-0002-8553-3302]{A.~Carramiñana}
\affiliation{Instituto Nacional de Astrof\'{i}sica, Óptica y Electrónica, Puebla, Mexico}

\author[0000-0002-6144-9122]{S.~Casanova}
\affiliation{Institute of Nuclear Physics Polish Academy of Sciences, PL-31342 IFJ-PAN, Krakow, Poland}

\author{O.~Chaparro-Amaro}
\affiliation{Centro de Investigaci\'on en Computaci\'on, Instituto Polit\'ecnico Nacional, M\'exico City, M\'exico.}

\author{U.~Cotti}
\affiliation{Universidad Michoacana de San Nicolás de Hidalgo, Morelia, Mexico}

\author{J.~Cotzomi}
\affiliation{Facultad de Ciencias F\'{i}sico Matemáticas, Benemérita Universidad Autónoma de Puebla, Puebla, Mexico }

\author[0000-0001-9643-4134]{E.~De la Fuente}
\affiliation{Departamento de F\'{i}sica, Centro Universitario de Ciencias Exactase Ingenierias, Universidad de Guadalajara, Guadalajara, Mexico }

\author{R.~Diaz Hernandez}
\affiliation{Instituto Nacional de Astrof\'{i}sica, Óptica y Electrónica, Puebla, Mexico}

\author[0000-0001-8451-7450]{B.L.~Dingus}
\affiliation{Physics Division, Los Alamos National Laboratory, Los Alamos, NM, USA }

\author[0000-0002-2987-9691]{M.A.~DuVernois}
\affiliation{Department of Physics, University of Wisconsin-Madison, Madison, WI, USA }

\author[0000-0003-2169-0306]{M.~Durocher}
\affiliation{Physics Division, Los Alamos National Laboratory, Los Alamos, NM, USA }

\author[0000-0002-0087-0693]{J.C.~Díaz-Vélez}
\affiliation{Departamento de F\'{i}sica, Centro Universitario de Ciencias Exactase Ingenierias, Universidad de Guadalajara, Guadalajara, Mexico }

\author[0000-0001-7074-1726]{C.~Espinoza}
\affiliation{Instituto de F\'{i}sica, Universidad Nacional Autónoma de México, Ciudad de Mexico, Mexico}

\author{K.L.~Fan}
\affiliation{Department of Physics, University of Maryland, College Park, MD, USA }

\author[0000-0002-0173-6453]{N.~Fraija}
\affiliation{Instituto de Astronom\'{i}a, Universidad Nacional Autónoma de México, Ciudad de Mexico, Mexico}

\author[0000-0002-4188-5584]{J.A.~García-González}
\affiliation{Tecnologico de Monterrey, Escuela de Ingenier\'{i}a y Ciencias, Ave. Eugenio Garza Sada 2501, Monterrey, N.L., Mexico, 64849}

\author[0000-0003-1122-4168]{F.~Garfias}
\affiliation{Instituto de Astronom\'{i}a, Universidad Nacional Autónoma de México, Ciudad de Mexico, Mexico}

\author[0000-0002-5209-5641]{M.M.~González}
\affiliation{Instituto de Astronom\'{i}a, Universidad Nacional Autónoma de México, Ciudad de Mexico, Mexico}

\author[0000-0001-9844-2648]{J.P.~Harding}
\affiliation{Physics Division, Los Alamos National Laboratory, Los Alamos, NM, USA }

\author[0000-0002-2565-8365]{S.~Hern\'andez-Cadena}
\affiliation{Instituto de F\'{i}sica, Universidad Nacional Autónoma de México, Ciudad de Mexico, Mexico}

\author[0000-0002-3808-4639]{D.~Huang}
\affiliation{Department of Physics, Michigan Technological University, Houghton, MI, USA}

\author[0000-0002-5527-7141]{F.~Hueyotl-Zahuantitla}
\affiliation{Universidad Autónoma de Chiapas, Tuxtla Gutiérrez, Chiapas, México}

\author[0000-0001-5811-5167]{A.~Iriarte}
\affiliation{Instituto de Astronom\'{i}a, Universidad Nacional Autónoma de México, Ciudad de Mexico, Mexico}

\author[0000-0003-4467-3621]{V.~Joshi}
\affiliation{Erlangen Centre for Astroparticle Physics, Friedrich-Alexander-Universit\"at Erlangen-N\"urnberg, Erlangen, Germany}

\author{S.~Kaufmann}
\affiliation{Universidad Politecnica de Pachuca, Pachuca, Hgo, Mexico }

\author[0000-0003-4785-0101]{D.~Kieda}
\affiliation{Department of Physics and Astronomy, University of Utah, Salt Lake City, UT, USA }

\author{J.~Lee}
\affiliation{University of Seoul, Seoul, Rep. of Korea}

\author[0000-0001-5516-4975]{H.~León Vargas}
\affiliation{Instituto de F\'{i}sica, Universidad Nacional Autónoma de México, Ciudad de Mexico, Mexico}

\author{J.T.~Linnemann}
\affiliation{Department of Physics and Astronomy, Michigan State University, East Lansing, MI, USA }

\author[0000-0001-8825-3624]{A.L.~Longinotti}
\affiliation{Instituto de Astronom\'{i}a, Universidad Nacional Autónoma de México, Ciudad de Mexico, Mexico}

\author[0000-0003-2810-4867]{G.~Luis-Raya}
\affiliation{Universidad Politecnica de Pachuca, Pachuca, Hgo, Mexico }

\author[0000-0001-8088-400X]{K.~Malone}
\affiliation{Space Science and Applications Group, Los Alamos National Laboratory, Los Alamos, NM USA}

\author[0000-0001-9052-856X]{O.~Martinez}
\affiliation{Facultad de Ciencias F\'{i}sico Matemáticas, Benemérita Universidad Autónoma de Puebla, Puebla, Mexico }

\author[0000-0002-2824-3544]{J.~Martínez-Castro}
\affiliation{Centro de Investigaci\'on en Computaci\'on, Instituto Polit\'ecnico Nacional, M\'exico City, M\'exico.}

\author[0000-0002-2610-863X]{J.A.~Matthews}
\affiliation{Dept of Physics and Astronomy, University of New Mexico, Albuquerque, NM, USA }

\author[0000-0002-1114-2640]{E.~Moreno}
\affiliation{Facultad de Ciencias F\'{i}sico Matemáticas, Benemérita Universidad Autónoma de Puebla, Puebla, Mexico }

\author[0000-0002-7675-4656]{M.~Mostafá}
\affiliation{Department of Physics, Pennsylvania State University, University Park, PA, USA}

\author[0000-0003-0587-4324]{A.~Nayerhoda}
\affiliation{Institute of Nuclear Physics Polish Academy of Sciences, PL-31342 IFJ-PAN, Krakow, Poland}

\author[[0000-0003-1059-8731]{L.~Nellen}
\affiliation{Instituto de Ciencias Nucleares, Universidad Nacional Autónoma de Mexico, Ciudad de Mexico, Mexico }

\author[0000-0002-5448-7577]{N.~Omodei}
\affiliation{Department of Physics, Stanford University: Stanford, CA 94305–4060, USA}

\author[0000-0002-8774-8147]{Y.~Pérez Araujo}
\affiliation{Instituto de Astronom\'{i}a, Universidad Nacional Autónoma de México, Ciudad de Mexico, Mexico}

\author[0000-0001-5998-4938]{E.G.~Pérez-Pérez}
\affiliation{Universidad Politecnica de Pachuca, Pachuca, Hgo, Mexico }

\author[0000-0002-6524-9769]{C.D.~Rho}
\affiliation{University of Seoul, Seoul, Rep. of Korea}

\author[0000-0003-1327-0838]{D.~Rosa-González}
\affiliation{Instituto Nacional de Astrof\'{i}sica, Óptica y Electrónica, Puebla, Mexico}

\author[0000-0003-4556-7302]{H.~Salazar}
\affiliation{Facultad de Ciencias F\'{i}sico Matemáticas, Benemérita Universidad Autónoma de Puebla, Puebla, Mexico }

\author[0000-0002-9312-9684]{D.~Salazar-Gallegos}
\affiliation{Department of Physics and Astronomy, Michigan State University, East Lansing, MI, USA }

\author[0000-0001-6079-2722]{A.~Sandoval}
\affiliation{Instituto de F\'{i}sica, Universidad Nacional Autónoma de México, Ciudad de Mexico, Mexico}

\author{J.~Serna-Franco}
\affiliation{Instituto de F\'{i}sica, Universidad Nacional Autónoma de México, Ciudad de Mexico, Mexico}

\author{Y.~Son}
\affiliation{University of Seoul, Seoul, Rep. of Korea}

\author[0000-0002-1492-0380]{R.W.~Springer}
\affiliation{Department of Physics and Astronomy, University of Utah, Salt Lake City, UT, USA }

\author[0000-0002-9074-0584]{O.~Tibolla}
\affiliation{Universidad Politecnica de Pachuca, Pachuca, Hgo, Mexico }

\author[0000-0001-9725-1479]{K.~Tollefson}
\affiliation{Department of Physics and Astronomy, Michigan State University, East Lansing, MI, USA }

\author[0000-0002-1689-3945]{I.~Torres}
\affiliation{Instituto Nacional de Astrof\'{i}sica, Óptica y Electrónica, Puebla, Mexico}

\author[0000-0002-7102-3352]{R.~Torres-Escobedo}
\affiliation{Tsung-Dao Lee Institute \& School of Physics and Astronomy, Shanghai Jiao Tong University}

\author[0000-0003-1068-6707]{R.~Turner}
\affiliation{Department of Physics, Michigan Technological University, Houghton, MI, USA}

\author[0000-0002-2748-2527]{F.~Ureña-Mena}
\affiliation{Instituto Nacional de Astrof\'{i}sica, Óptica y Electrónica, Puebla, Mexico}

\author[0000-0001-6876-2800]{L.~Villaseñor}
\affiliation{Facultad de Ciencias F\'{i}sico Matemáticas, Benem\'erita Universidad Aut\'onoma de Puebla, Puebla, Mexico }

\author{X.~Wang}
\affiliation{Department of Physics, Michigan Technological University, Houghton, MI, USA}

\author[0000-0002-6623-0277]{E.~Willox}
\affiliation{Department of Physics, University of Maryland, College Park, MD, USA }

\author[0000-0003-0513-3841]{H.~Zhou}
\affiliation{Tsung-Dao Lee Institute \& School of Physics and Astronomy, Shanghai Jiao Tong University}

\author[0000-0002-8528-9573]{C.~de León}
\affiliation{Universidad Michoacana de San Nicolás de Hidalgo, Morelia, Mexico}

\collaboration{70}{The HAWC Collaboration}

\author[0000-0003-1826-6117]{V.~Gammaldi}
\affiliation{Departamento de F\'isica Te\'orica, Universidad Aut\'onoma Madrid}
\affiliation{Instituto de F\'isica Te\'orica, IFT UAM-CSIC}

\author[0000-0001-8260-4147]{E.~Karukes}
\affiliation{AstroCeNT, Nicolaus Copernicus Astronomical Center Polish Academy of Sciences, ul. Rektorska 4, 00-614 Warsaw, Poland}

\author[0000-0002-5476-2954]{P.~Salucci}
\affiliation{SISSA, International School for Advanced Studies, Via Bonomea 265, 34136, Trieste, Italy}
\affiliation{INFN, Istituto Nazionale di Fisica Nucleare - Sezione di Trieste, Via Valerio 2, 34127, Trieste, Italy}



\begin{abstract}
We present the results of dark matter (DM) searches in a sample of 31 dwarf irregular (dIrr) galaxies within the field of view of the HAWC Observatory. dIrr galaxies are DM dominated objects, which astrophysical gamma-ray emission is estimated to be negligible with respect to the secondary gamma-ray flux expected by annihilation or decay of Weakly Interacting Massive Particles (WIMPs). While we do not see any statistically significant DM signal in dIrr galaxies, we present the exclusion limits ($95\%~\text{C.L.}$) for annihilation cross-section and decay lifetime for WIMP candidates with masses between $1$ and $100~\text{TeV}$. Exclusion limits from dIrr galaxies are relevant and complementary to benchmark dwarf Spheroidal (dSph) galaxies. In fact, dIrr galaxies are targets kinematically different from benchmark dSph, preserving the footprints of different evolution histories. We compare the limits from dIrr galaxies to those from ultrafaint and classical dSph galaxies previously observed with HAWC. We find that the contraints are comparable to the limits from classical dSph galaxies and $\thicksim2$ orders of magnitude weaker than the ultrafaint dSph limits.

\end{abstract}



\section{Introduction}
\label{sec:intro}

While we have clear evidence of the existence of dark matter (DM) in the Universe, we do not know the nature of DM, and various candidates have been proposed. One of the most common DM candidates are the Weakly Interacting Massive Particles (WIMPs), a family of particles with interaction at the weak scale that can be thermally produced in the early universe. More details can be found in \cite{bertone2005,bertone2018,ARBEY2021103865}. We assume that the generic WIMP (as the neutralino predicted by SUSY) is the only component of DM. From the direct searches and collider experiments (see \cite{ARBEY2021103865,Schumann_2019,Mitsou_2015} and references therein), very restrictive constraints have been established in the mass range from $5~\text{GeV}$ to  $1 ~\text{TeV}$. Indirect searches explore a broad range of masses, from 10 GeV up to the unitarity limit ($400~\text{TeV}$ for a Majorana particle \citep{ENQVIST1991367}, or $144~\text{TeV}$ considering the effect of bound states \citep{smirnov_2019}) by assuming that the DM particles can annihilate or decay to Standard Model (SM) particles producing stable final-states as protons, neutrinos and gamma-rays. Then, indirect DM searches allow us to explore  candidates with masses (above $10~\text{TeV}$) that are not accessible to LHC or direct detection experiments. Indeed, by using observational data of wide-field gamma-ray observatories we can constrain the range of fundamental parameters of DM candidates. To do this, we use DM dominated astrophysical objects where the probability to observe a DM signal is high. The target with the highest expected DM signal in the gamma-ray sky is the galactic center, but the analysis of this region involves the understanding of the distribution and amount of astrophysical gamma-ray sources in the region around the center. We can also observe regions where we expect negligible gamma-ray emission from astrophysical processes, which would lead to a clean DM signal. Example of these objects are not only very well-know dwarf spheroidal (dSph), but also dwarf irregular (dIrr) galaxies. The dIrr galaxy's population is characterized by very low star-formation rates (SFR) \citep{McGaugh2017} and the lack of massive stars \citep{regina1998,dunn2007}. Hence there is little background gamma-ray emission at energies above $1~\text{TeV}$ and dIrr can be considered as essentially background-free objects in indirect DM searches (see e.g. \cite{dirrDMTheoretical}). Here, we use a sample of dIrr galaxies within the High Altitude Water Cherenkov (HAWC) field-of-view to show the HAWC sensitivity to look for DM signatures in these objects, compared with the limit obtained by benchmark targets \citep{hawcdsph}. The paper is organized as follows: in section \ref{hawc} we briefly introduce the HAWC Observatory. In section \ref{dirrs} we present the sample of dIrr galaxies used for this study and discuss the DM density profiles computed for dIrrs. In section \ref{DMana} we discuss the data set and the analysis for individual sources, section \ref{SSana}, and the combined analysis for this population, section \ref{JLana}. In section \ref{results} we show the exclusion limits computed for 31 dIrr galaxies and the comparison with limits from dSph galaxies. Finally, we conclude in section \ref{conclusions}.

\section{The HAWC Observatory}\label{hawc}

Situated in Sierra Negra, Mexico at an altitude of $4100~\text{m}$, the HAWC Observatory is an extended array of $300$ Water Cherenkov Detectors (WCDs) to detect air showers produced by very high energy (VHE) gamma rays and cosmic rays. Every WCD is $7.3~\text{m}$ in diameter and $4.5~\text{m}$ deep, filled with $200000$ liters of purified water and instrumented with four Photo Multiplier Tubes (PMTs) anchored to the bottom. The PMTs collect the Cherenkov light produced by charged particles passing through the WCDs. One of the PMTs is located in the middle, and the other three are located at a distance of $1.8~\text{m}$ from the central PMT and equally spaced ($120^{\circ}$). The HAWC Observatory is sensitive to gamma rays with energies in the range between $1~\text{TeV}$ and $100~\text{TeV}$, with a wide field of view, covering $2/3$ of the sky each day and a duty cycle $>95\%$. This allows the HAWC Observatory to study large populations of astrophysical sources and constrain parameters that are common to all the targets in the sample.\\% In this job, we apply this capability to set a $95\%$ C.L. exclusion limits in annihilation cross section and decay lifetime of DM particles for a local population of dwarf Irregular (dIrr) galaxies.\\
When an air shower reaches the detector, hits recorded during the air shower event are selected within a time window of $250~\text{ns}$ around the trigger time. Hits that survive the time selection and quality cuts are used for reconstruction of the air shower parameters. After reconstruction, analysis of air shower events is based on cuts depending on the energy and size of events measured in the detector. Then, HAWC events are organized in bins according to the size of the area covered by each event recorded on the ground. The event size is defined as the ratio of the number of PMT hits used for reconstruction to the total number of available PMTs for reconstruction. A range of values of this ratio is called a fractional bin. Table \ref{bins} shows the definition of the 9 fractional bins, $f_{\text{hit}}$ used in HAWC. This definition of fractional bins according to the size of the air shower recorded in the HAWC Observatory is weakly correlated to the energy of the primary particle, \citep{Abeysekara_2017crab,Abeysekara_2019}. For more details about the reconstruction and estimation of physical parameters of the air showers in HAWC see \cite{Abeysekara_2017,Abeysekara_2017crab,Albert_2020}.

\begin{table}[htb!]
\centering
\begin{tabular}{c||c|c}
Bin & $f_{\text{hit}}$ & \textbf{Angular Resolution} \\ 
 -- & -- & \textbf{(deg)}\\\hline
1 & 6.7\% \;\,-- 10.5\% & \textbf{0.98}\\
2 & 10.5\% -- 16.2\% & \textbf{0.71} \\
3 & 16.2\% -- 24.7\% & \textbf{0.53}\\
4 & 24.7\% -- 35.6\% & \textbf{0.39}\\
5 & 35.6\% -- 48.5\% & \textbf{0.33}\\
6 & 48.5\% -- 61.8\% & \textbf{0.30}\\
7 & 61.8\% -- 74.0\% & \textbf{0.24}\\
8 & 74.0\% -- 84.0\% & \textbf{0.21}\\
9 & \;\;84.0\% -- 100.0\% & \textbf{0.16}\\
\end{tabular}
\caption{Definition of fractional $f_{\text{hit}}$ bins used in HAWC. The second column shows the range of values of the PMTs used for reconstruction, and the third column shows the angular resolution measured from observation of the Crab nebula \citep{Abeysekara_2017crab}}
\label{bins}
\end{table}

\begin{sidewaysfigure}
\centering
\includegraphics[width=\linewidth]{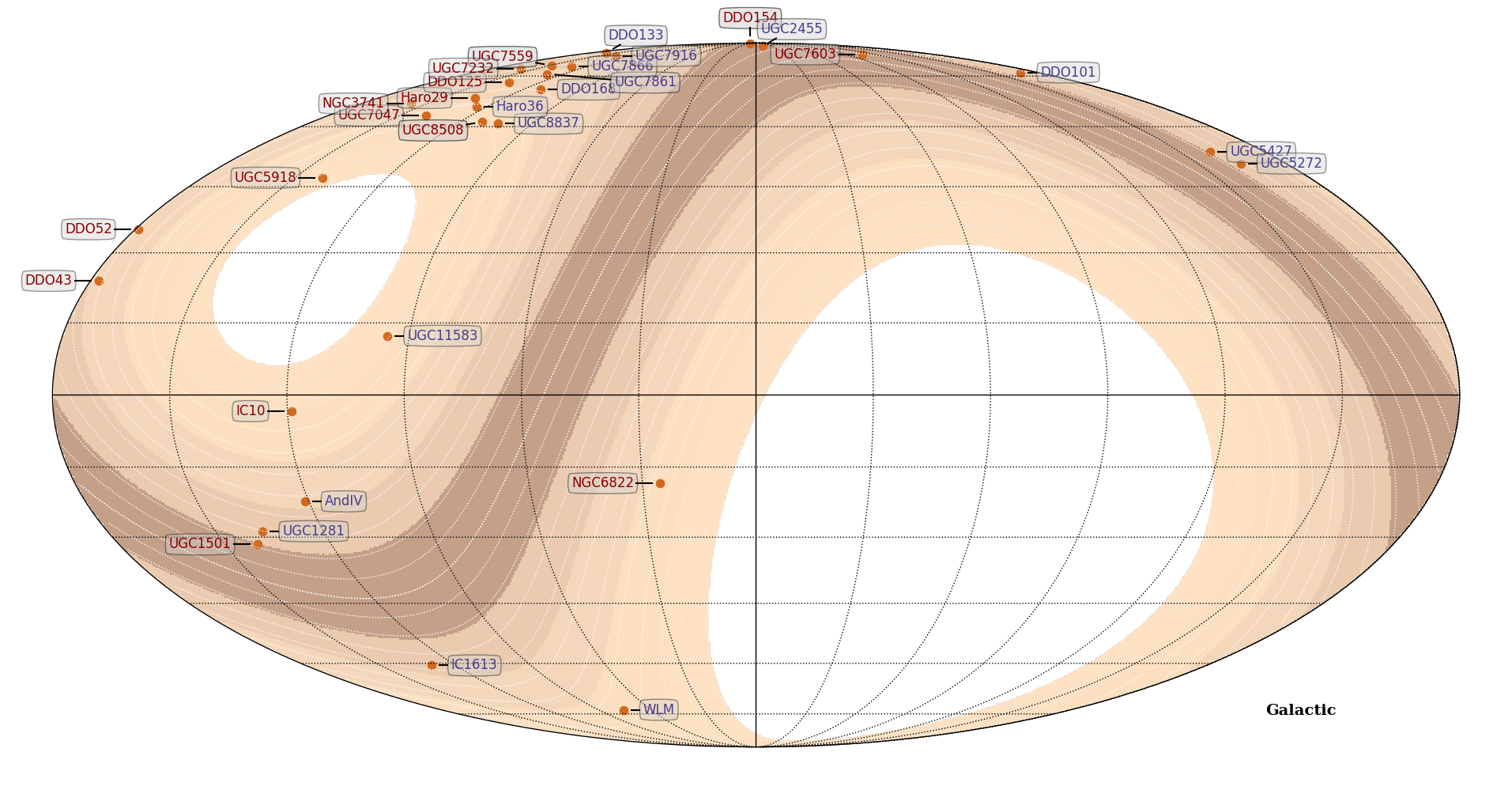}
\caption{Sky Map. The color bands show different HAWC sensitivities from most sensitive (dark brown) to least sensitive (light brown). The white regions are outside of the field of view of HAWC. The map shows the position of all the dIrr galaxies considered in this study.}
\label{skymap}
\end{sidewaysfigure}

\section{Dark Matter Models for dIrr Galaxies}\label{dirrs}

Figure \ref{skymap} shows the sky map and position of our sample of 31 dIrrs in the HAWC field of view. All the dIrr in the sample have a distance to Earth smaller than $11~\text{Mpc}$. All of these galaxies are extragalactic sources, with some of them belonging to the Local Group of galaxies; and redshifts $z<0.001$. The sample is taken from \cite{karukes2016}, where the authors computed the DM properties of 36 dIrr Galaxies using kinematical data, and after select the galaxies within the HAWC field of view, the sample is reduced to 31 dIrr galaxies. They establish that these galaxies follow a low-mass version of the Universal Rotation Curve (URC) function.  The URC is a generic function of radial distance to the center of a spiral galaxy that describes the rotation curve of all galaxies in the local volume, and parameterizes the distributions of matter inside a galaxy \citep{urc1996}. These parameters are found by fitting three contributions to the rotational motion in the galaxy: a stellar disk, a HI disk and a spherical DM halo. In the case of DM halo, the profile with the best fit is constrained to be a Burkert density profile \citep{Burkert1995}:

\begin{equation}
\rho_{\text{DM}}(r)=\rho_0\frac{r_0^3}{(r_0+r)(r_0^2+r^2)}
\end{equation}
\noindent
where $r_0$ is the core radius, and $\rho_0$ is the normalization density. These observational core profiles represent a modified, isothermal sphere which better fit the observations of the rotation curves that cannot be explained by DM cusped profiles, resulting from N-body cosmological simulations \citep{nfw1,fermidIrrs}. The 
$\rho_0$ and $r_0$ parameters used in this study are taken from \cite{karukes2016}. Table \ref{dIrrSample} shows the sample of galaxies, their coordinates, virial size and the value of astrophysical factors computed for these galaxies.

\begin{table}[htb!]
\centering
\begin{tabular}{l||c|c|c|c|c}
Name & R.A. & DEC. & $\theta_{\text{vir}}$ & $\log_{10}(\frac{\mathrm{J}}{\text{TeV}^2\text{cm}^5})$ & $\log_{10}(\frac{\mathrm{D}}{\text{TeV}\text{cm}^2})$  \\
& (deg) & (deg) & (deg) & & \\ \hline\hline
And IV &10.62 & 40.57 & 0.326 & 9.764 & 13.463 \\
DDO 101 & 177.91 & 31.51 & 0.309 & 10.356 & 14.312 \\
DDO 125 & 186.92 & 43.493 & 0.533 & 10.467 & 14.165 \\
DDO 133	& 188.22 & 31.54 & 0.784 & 11.501 & 15.274 \\
DDO 154	& 193.52 & 27.15 & 0.695 & 11.800 & 15.397 \\
DDO 168 & 198.61 & 45.91 & 1.142 & 11.365 & 15.271 \\
DDO 43 & 112.07 & 40.77 & 0.472 & 10.109 & 13.853 \\
DDO 52 & 127.11 & 41.85 & 0.570 & 10.452 & 14.401 \\
Haro 29 & 186.56 & 48.49 & 0.333 & 9.974 & 13.764 \\
Haro 36 & 191.73 & 51.61 & 0.550 & 10.642 & 13.581 \\
IC 10 & 5.10 & 59.29 & 3.857 & 11.857 & 15.619 \\
IC 1613 & 16.19 & 2.13 & 2.361 & 11.632 & 15.325 \\
NGC 3741 & 174.02 & 45.28 & 0.405 & 9.814 & 13.417 \\
NGC 6822 & 296.23 & -14.80 & 5.325 & 12.173 & 15.943 \\
UGC 11583 & 307.56 & 60.44 & 0.853 & 10.676 & 14.605 \\
UGC 1281 & 27.38 & 32.59 & 0.952 & 10.854 & 14.739 \\
UGC 1501 & 30.31 & 28.84 & 1.032 & 10.937 & 14.843 \\
UGC 2455 & 194.92 & 25.23 & 0.569 & 10.392 & 14.250 \\
UGC 5272 & 147.59 & 31.48 & 0.765 & 10.721 & 14.731 \\
UGC 5427 & 151.17 & 29.36 & 0.436 & 10.133 & 14.007 \\
UGC 5918 & 162.40 & 65.53 & 0.612 & 10.512 & 14.420 \\
UGC 7047 & 181.01 & 52.58 & 0.649 & 10.630 & 14.444 \\
UGC 7232 & 183.43 & 36.63 & 0.654 & 10.845 & 14.581 \\
UGC 7559 & 186.77 & 37.14 & 0.700 & 11.105 & 14.938 \\
UGC 7603 & 187.18 & 22.82 & 0.652 & 11.368 & 15.251 \\
UGC 7861 & 190.46 & 41.27 & 0.535 & 10.804 & 14.715 \\
UGC 7866 & 190.56 & 38.50 & 0.496 & 10.672 & 14.462 \\
UGC 7916 & 191.10 & 34.38 & 0.489 & 11.012 & 14.898 \\
UGC 8508 & 202.68 & 54.91 & 0.584 & 10.362 & 14.071 \\
UGC 8837 & 208.68 & 53.90 & 0.737 & 10.856 & 14.802 \\
WLM & 0.49 & -15.46 & 2.609 & 12.062 & 15.777 \\
\end{tabular}
\caption{Sample of dIrr galaxies. We show the 31 dIrr galaxies within the HAWC field-of-view used in this study. Columns: name of the galaxy (1), the right ascension ($\alpha$) (2) and declination ($\delta$) (3) of the galaxy, the angular extension of the virial radius (4), the astrophysical factor for annihilation (5) and decay (6), computed with \textsc{Clumpy} \citep{clumpyv3}}
\label{dIrrSample}
\end{table}

\subsection{DM photon flux}\label{flux}

%Although there are several candidates to explain the DM content in the Universe, we assume that the generic WIMP (as the neutralino predicted by SUSY) is the only component of DM. From the direct searches and collider experiments (see \cite{Schumann_2019,Mitsou_2015} and references therein), very restrictive constraints have been established in the mass range from $5~\text{GeV}$ to  $1 ~\text{TeV}$. Indirect searches explore a broad range of masses, from 10 GeV up to the unitarity limit ($400~\text{TeV}$ for a Majorana particle \cite{ENQVIST1991367}, or $144~\text{TeV}$ considering the effect of bound states \cite{smirnov_2019}). In particular the energy sensitivity of the HAWC Observatory allow us to investigate WIMPs with masses form 1 to 100 TeV. We assume that DM particles can annihilate or decay to Standard Model particles producing stable final-states as protons, neutrinos and gamma rays. We are interested in estimating the flux of gamma-ray photons produced during annihilation or decay of the DM particles.

% \begin{equation*}
% \frac{\mathrm{d}\Phi_{\gamma}^{\text{DM}}}{\mathrm{d}E}\,=\,\text{P.P.} \times \text{A.F.}(\rho_{\text{DM}})
% \end{equation*}
\noindent
% The first term is called Particle Physics Factor and describes the nature of the DM candidate and the number of photons produced per energy bin in every DM annihilation or decay. The second term, called Astrophysical Factor, is associated to the geometrical properties of the DM content in the object of interest.

In this section we will consider both annihilation and decay of WIMPs. In the case of annihilation, the differential flux of gamma-ray photons is:

\begin{equation}
\frac{\mathrm{d}\Phi_{\gamma}^{\text{ann}}}{\mathrm{d}E}\,=\,\frac{\langle\sigma_{\chi} v\rangle^{\text{ref}}}{8\pi m_{\chi}^2}\sum_f B_f\frac{\mathrm{d}N_{f}^{\text{ann}}}{\mathrm{d}E}\times \int_{\Delta\Omega}\int_{l.o.s.}\,\mathrm{d}l\mathrm{d}\Omega\,\rho_{\text{DM}}\left(r(l)\right)^{2}
\label{dmFluxAnn}
\end{equation}
\noindent
where $\mathrm{d}N_{f}^{\text{ann}}/\mathrm{d}E$ is the differential spectrum of photons produced for annihilation channel $f$, $m_{\chi}$ is the mass of the WIMP, $\langle\sigma_{\chi}v\rangle^{\text{ref}}$ is the reference thermal averaged annihilation cross-section. The sum is over the total number of channels with branching ratios $B_f$ that contribute to the production of photons ($\sum_f{B_f}=1$). We assume that branching ratios have a value of $1$ for a specified channel, while the others have $B_f\,=\,0$. The term in the integral is called the Astrophysical Factor J, or J factor. The J factor is the double integral of the DM density profile squared along the line of sight $l$ and over the solid angle $\Delta\Omega$ around the line of sight.

For decaying DM, the differential flux of photons is computed by:

\begin{equation}
\frac{\mathrm{d}\Phi_{\gamma}^{\text{dec}}}{\mathrm{d}E}\,=\,\frac{1}{4\pi m_{\chi}\tau_{\chi}^{\text{ref}}}\sum_f B_f\frac{\mathrm{d}N_{f}^{\text{dec}}}{\mathrm{d}E}\times \int_{\Delta\Omega}\int_{l.o.s.}\,\mathrm{d}l\mathrm{d}\Omega\,\rho_{\text{DM}}\left(r(l)\right)
\label{dmFluxDec}
\end{equation}
\noindent
As in the case of annihilation, $\mathrm{d}N_{f}^{\text{dec}}/\mathrm{d}E$ is the differential spectrum of photons produced for a decay channel $f$, $\tau_{\chi}^{\text{ref}}$ is a reference lifetime of the DM candidate. For DM decay, the Astrophysical Factor D or D-factor, is the double integral of the DM density profile $\rho_{\text{DM}}$ along the line of sight $l$ and over the solid angle $\Delta\Omega$ around the line of sight. We show the sample of galaxies and their values of J- and D-factors in Table \ref{dIrrSample}. 

\subsubsection{Photon Spectra}
\label{spectre}

\begin{figure}[ht!]
%\captionsetup{width=0.8\linewidth}
\centering
\includegraphics[width=0.9\linewidth]{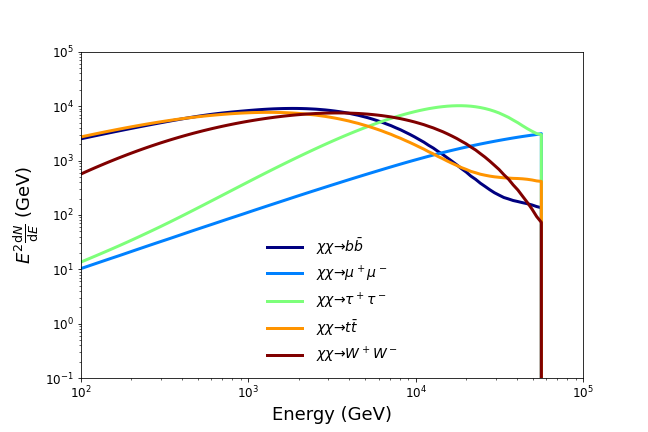}
\caption{Spectrum of Photons computed for annihilation of WIMPs with mass of 60 TeV to five channels, assuming that the branching ratio for each channel is $100\%.$}
\label{spectra}
\end{figure}

The production of photons from annihilation (decay) of DM particles is due to the decay or hadronization processes of the unstable products. In both cases, the spectrum of photons is continuous and has an energy cutoff at the energy available in the process, the mass  (half-mass) of the DM particle for annihilation (decay). For this work, we considered WIMP masses in the range from $1~\text{TeV}$ to $100~\text{TeV}$ and annihilation (decay) to five channels: $b$ and $t$ quarks, $\mu$ and $\tau$ leptons, and the $W$ boson. The spectrum of photons is calculated with \textsc{Pythia 8} \citep{pythia}. Figure \ref{spectra} shows the spectrum of photons for the annihilation of a WIMP with mass $m_{\chi}=60$ TeV.

\subsubsection{The Astrophysical Factor}

For the set of dIrr Galaxies we studied here, the structural properties of luminous and DM contributions are constrained using kinematical data taken from \cite{karukes2016}. The DM density is described by a Burkert profile \citep{Burkert1995}. The Burkert profile is a density distribution that resembles an isothermal profile in the inner regions ($r<r_0$) and a distribution with slope $-3$ in the outer regions. For this study, the virial radius, $R_{\text{vir}}$ and other DM related parameters were computed assuming an overdensity $\Delta$ equal to $200$, and a $M_{DM}$--$c_{\text{vir}}$ taken from \cite{masc2014}. We do not take into account the presence of DM sub halos as estimates of the boost factor may increase the total gamma-ray flux up to $\thicksim10\times$ the contribution of the main halo. As we assume a point-source model for the DM-induced gamma-ray emission; the total boost factor is traduced as a multiplicative factor to the exclusion limits. However, this boost factor does not change the conclusions of the analysis.
%Because of the angular extension $\theta_\text{vir}$ of the DM haloes (column 4 in Table \ref{dIrrSample}), we do not consider any enhancement due to subhaloes in the main DM halo.
Galaxies with $\theta_{\text{vir}}>1^{\circ}$ are left for a later study, and we assume that the spatial emission from the galaxies is coming from a point source as the angular resolution for bins 1 and 2 is $0.98~\text{deg}$ and $0.71~\text{deg}$ respectively, see Table \ref{dIrrSample}. In order to compute the integrals for J and D factors, we use the \textsc{Clumpy} package \citep{clumpyv3}. We compute the astrophysical factors over the total extension of the DM halo.

In Table \ref{dIrrSample}, we report the angular virial radius $\theta_{\text{vir}}$ (column 4), the J and D factors (columns 5 and 6) for dIrrs. According to these values, we observe that the best targets in the sample are the galaxies NGC 6822, IC 10, IC 1613, WLM and DDO 154 because of their large J and D factors. In particular, the galaxy DDO 154 is located in a declination band that is favorable to the HAWC Observatory, see Figure \ref{skymap}. Furthermore, the values of D and J factors in our work are comparable to the values reported in \cite{dirrDMTheoretical,fermidIrrs}. In particular, the slightly difference with respect to the values in \cite{dirrDMTheoretical} is because the authors assume a value of overdensity, $\Delta$, equal to $100$. Differences with respect to the values reported in \cite{fermidIrrs} is due to the reanalysis of availabale kinematic data to obtain the parameters of the DM profile. We also note, that the best targets in the sample reported here are consistent with the results obtained in \cite{dirrDMTheoretical,fermidIrrs}.

Additionally, we can compare the astrophysical factors to the values for dSph galaxies, in particular with galaxies within the field of view of the HAWC Observatory. We used the sample of galaxies studied in \cite{hawcdsph}. As we describe later (see Section \ref{comparison}), the population of dSph can be divided into two subclasses according to the number of stars hosted in the galaxies: classical and ultrafaint galaxies. The values of the J and D factors for dSph galaxies are shown in Table \ref{dsphsample}. We can observe that values of J factors for dIrr galaxies are roughly between two and three orders of magnitude smaller than the values for ultrafaint galaxies, while this difference is reduced for classical dSph galaxies. This may indicate that dIrr galaxies are not very suited to perform searches of annihilating dark matter. However, one should consider that values of J factors for ultrafaint dSphs, in the majority of cases, have larger uncertainties due to the lack of stellar data to constrain the dark matter profile. The case for decay is more interesting, as we observe that the value of D factors for the three different populations are very similar, so we will consider combined analyses between dIrr and dSph populations as a future work.

\begin{table}[htb!]
\centering
\begin{tabular}{l||c|c}
Name & $\log_{10}(\frac{\mathrm{J}}{\text{TeV}^2\text{cm}^5})$ & $\log_{10}(\frac{\mathrm{D}}{\text{TeV}\text{cm}^2})$  \\
&&\\
\hline \hline
\multicolumn{3}{c}{Ultrafaint}\\
\hline
Triangulum II & 14.44 & 15.42\\
Segue I & 13.66 & 15.64 \\
Coma Berenices & 13.32 & 15.71 \\
\hline
\multicolumn{3}{c}{Classical}\\
\hline
Draco & 13.37 & 16.15 \\
Leo I & 11.57 & 15.04 \\
Leo II & 12.11 & 14.33 \\
Ursa Minor & 13.24 & 14.92
\end{tabular}
\caption{Values of J and D factors for dSph galaxies observed with HAWC. Values are taken from \cite{hawcdsph}. We convert the values to the same units for J and D factors for dIrr galaxies reported in Table \ref{dIrrSample}}
\label{dsphsample}
\end{table}

\begin{figure}[ht!]
    \centering
    \includegraphics[width=0.9\linewidth]{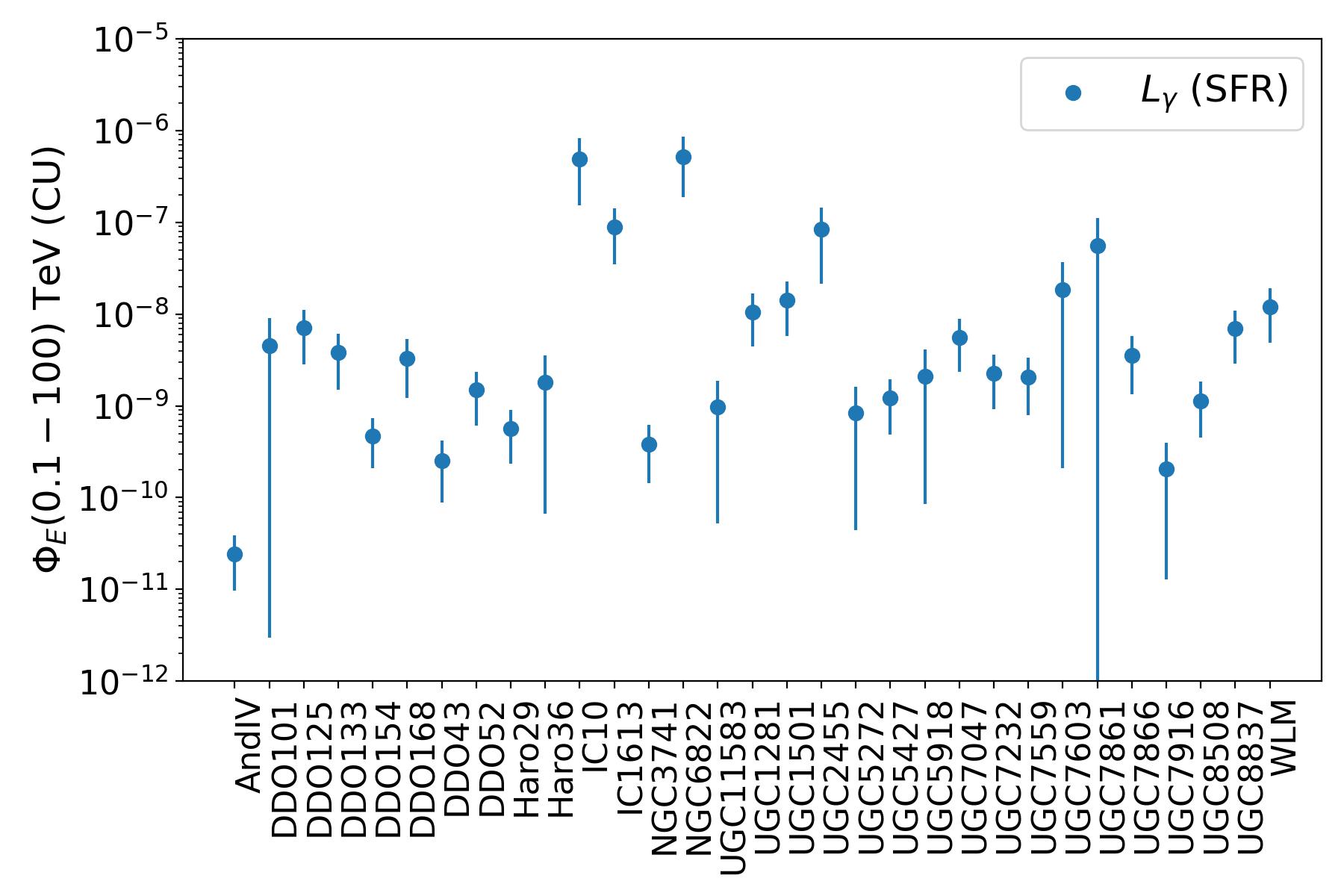}
    \caption{Expected Integral flux for dIrr galaxies in our sample. The gamma-ray luminosity is estimated using the SFR computed from the stellar mass for every galaxy. The integral flux is computed for energies between $100~\text{GeV}$ and $100~\text{TeV}$ and converted to HAWC Crab units (CU). See the text for more details.}
    \label{fig:sfr}
\end{figure}

\subsection{The Astrophysical gamma-ray emission}

Galaxies that present star-formation regions have been reported as gamma-ray emitters \citep{hesshepms}. Because of this, it is necessary to consider this emission to compute the photon fluxes. However, the galaxies we use in this study are characterized by very-low star formation rates \citep{dunn2007,regina1998,McGaugh2017}. Therefore gamma-ray emission to TeV energies is considered negligible. We use the method described in \cite{McGaugh2017} to estimate the SFR of the galaxies in our sample. Then, following the method described in \cite{martinsfr}, we estimate the gamma-ray flux of the dIrr galaxies to be $\lesssim7$ orders of magnitude below the Crab flux (see Figure \ref{fig:sfr}). The energy-integrated sensitivty to a point source with spectral index of $\alpha=-2.5$ in the position of the Crab Nebula is $0.028~\text{Crab Units (CU)}$, see Figure 2 in \cite{Albert_2020}, so we do not consider the SFR-induced gamma-ray emission in our gamma-ray model.

\section{Data and Analysis}\label{DMana}

To derive the exclusion limits, we use data from the HAWC Observatory comprising 1017 days of data taking. We select this period of time to be able to compare with the results for dSph galaxies in \cite{hawcdsph}. %The ratio between the two periods of time, $r=t^{\text{dIrr}}/t^\text{dSph}$, leads to and overall improvement of the limits for dSph galaxies of $\sqrt{r}=1.42$. We discard this value as the discussion and conclusions do not change, and we compare directly the results obtained from both dIrr and dSph populations.

To test our DM model, we use the Maximum Likelihood Method to estimate best-fit values and exclusion limits for the parameters of interest. This method constrains the values of free parameters by maximizing the likelihood function given a data set $D$. For the HAWC Observatory, we used the binned log-likelihood function given by:

\begin{equation}
\log{L( S_{i,k} , B_{i,k} | D )} = \sum_{i,k}\left[N_{i,k}\log{(B_{i,k} + S_{i,k})} -  \log{(N_{i,k}!)} - ( B_{i,k} + S_{i,k} ) \right]
\end{equation}\label{sslikelihood}
\noindent
where  $S_{i,k}$ is the number of expected events from our DM model, $B_{i,k}$ is the number of observed background events in the region of interest (ROI) around the position of the galaxy; and $N_{i,k}$ is the total number of observed counts. The sum is performed over $i$ spatial bins and $k$ $f_{\text{hit}}$ fractional bins. The signal events $S_{i,k}$ were obtained from the convolution of the photon-flux of our DM model for five annihilation (decay) channels and the Response Matrix of the HAWC Observatory.

To measure how the DM model fits the data set $D$, we use the Test Statistic (TS) provided by the likelihood profile:

\begin{equation}
\text{TS} = -2\log{\left(\frac{L( S_{i,k} = 0 , B_{i,k} )}{L( \hat{S}_{i,k} , B_{i,k} )}\right)}
\end{equation}\label{ssts}
\noindent
The numerator is the maximum likelihood value for the null hypothesis, assuming $S_{i,k}$ equal to zero in our DM model and the denominator is the maximum likelihood value when the DM model has signal $\hat{S}_{i,k}$ different from zero. As usual, the statistical significance is obtained by $\sigma=\sqrt{TS}$. We performed two kinds of analysis: for every dIrr Galaxy (Individual Source Analysis) and combined analysis. We did not see any statistically significant excess from the data, and the significance is transformed into (one-sided) exclusion limits for the annihilation cross-section and the decay lifetime of a generic DM candidate with masses in the range $[1,100]~\text{TeV}$. 

\subsection{Individual Source Analysis}\label{SSana}

For the individual analysis, the values of the background and signal are estimated for spatial pixels within a circular ROI of $5^\circ$ in diameter. The number of expected events for every source is computed by convolving the DM photon flux for annihilation (decay) and the response matrix of HAWC. The DM photon flux is calculated assuming reference values for the annihilation cross-section $\langle\sigma_{\chi}v\rangle^{(\text{ref})}$  and the decay lifetime $\tau_{\chi}^{(\text{ref})}$ of DM particles. Then, the $-\log{L( S_{i,k} , B_{i,k} | D)}$ is computed and minimized using the \textsc{Minuit} package \citep{minuit2}.

In all the cases the observations are consistent with the null hypothesis (no photons produced by annihilation or decay of DM particles), and the significances were converted into upper (lower) exclusion limits for DM annihilation cross-section (decay lifetime). To estimate the (one-sided) exclusion Limit (at 95\% CL), we found the value of signal parameter $\xi$ where the log-likelihood ratio changes in $2.71$ with respect to the position of the maximum:

\begin{equation}
\left[ \text{TS}^{(\text{max})} - \text{TS}(\xi)\right] - 2.71 = 0
\end{equation}
\noindent
The signal parameter $\xi$ is a global scale factor that has effect only on the signal parameter. $\text{TS}(\xi)$ is computed by:

\begin{equation}
\text{TS}(\xi) = 2 \sum_{i,k}\left[\log{\left(\frac{B_{i,k} + \xi\times S_{i,k}}{B_{i,k}}\right)} - \xi\times S_{i,k}\right]
\end{equation}
\noindent
When the signal parameter $\xi$ is found, the value of the annihilation cross-section and decay lifetime are obtained by the multiplication of the scale-factor and the reference values used to compute the DM photon flux:

\begin{eqnarray}
\langle\sigma_{\chi}v\rangle^{(95\%)} & = \langle\sigma_{\chi}v\rangle^{(\text{ref})} \times \xi\\
\tau_{\chi}^{(95\%)} & = \tau_{\chi}^{(\text{ref})} \times \xi^{-1}
\end{eqnarray}
\noindent
The values obtained for $\langle\sigma_{\chi}v\rangle^{(95\%)}$ and $\tau_{\chi}^{(95\%)}$ do not depend in the value chosen for $\langle\sigma_{\chi}v\rangle^{(\text{ref})}$ and $\tau_{\chi}^{(\text{ref})}$, whose values are selected only for computational convenience. The methods described here are included in the analysis software for the HAWC Observatory as the \textsc{Liff} package \citep{liff}.

\subsection{Combined Analysis}\label{JLana}

We used the joint analysis technique to compute a combined limit for all the dIrr galaxies. The joint analysis allows us to estimate the value of parameters that are common to different sets of data; either observations of a source carried out by different experiments or observations of different sources performed by the same experiment. In our DM model, the common parameter for all the sources is the annihilation cross-section (decay lifetime) used to compute the expected events for a specific source. The joint analysis is based on the joint likelihood function that results from the multiplication of the likelihoods for every dataset. In the case of the joint analysis of a sample of astrophysical sources, the joint log-likelihood is:

\begin{align}
\log{L_{\text{joint}}( A\times S_{i,k,m} , B_{i,k,m} | D )} & = \sum_{m=1}^{M} \log{L_{d}( A\times S_{i,k,m},B_{i,k,m} | D )}\\
& =\sum_{m=1}^{M} \sum_{i,k}\left[N_{i,k,m}\log{(B_{i,k,m} + A\times S_{i,k,m})} -  \log{N_{i,k,m}!} - ( B_{i,k,m} + A\times S_{i,k,m}) \right]
\end{align}\label{joint_likelihood}
\noindent
where $S_{i,k,m}$ is the number of expected events for the $m$-th source, the $i$-th spatial bin and the $k$-th $f_{\text{hit}}$ fractional bin. The common parameter for all the dIrrs is represented by a global normalization factor $A$. $N_{i,k,m}$ and $B_{i,k,m}$ are the observed counts and the background counts for the $m$-th source, the $i$-th spatial bin and the $k$-th nHit fractional bin respectively. In this way, the joint likelihood is the natural method of taking into account the fluctuations of the background of every galaxy in the estimation of the signal parameter. Now, we can define the Test Statistic as usual:

\begin{equation}
\text{TS}_{\text{joint}} = -2\log{\left(\frac{L_{\text{joint}}( S = 0 , B_{i,k,m} )}{L_{\text{joint}}( \hat{S} , B_{i,k,m} )}\right)}
\end{equation}\label{TS_joint}

In the case of no detection, the statistical significances are converted into (one-sided) exclusion limits following the same method as in the individual source analysis. We use the Minuit package \citep{minuit2} to find the common signal that minimizes the negative of the joint likelihood function.

\subsection{Expected Limits}

The combined limit is expected to be better than the best individual limit, constraining the common parameter even more. This occurs when all the sources in the sample have similar properties and background counts. However, all the sources have different astrophysical factors and background counts can fluctuate statistically, so the combined limit is not necessarily better, and it can even be worse, than the exclusion limit obtained by the best individual target in the sample. For example, in \cite{BHdSph}, a sample of two galaxies where the first galaxy has twice the background rate and similar astrophysical values of the second galaxy; the combined limit is $10\%$ worse than the exclusion limit of the best target in the sample.

To account for this statistical variation, we compute the expected limits for annihilation cross-section and decay lifetime to compare with the actual results obtained for dIrr galaxies. The expected limits are obtained by fluctuating the background counts of HAWC maps and applying the same pipeline for individual analysis. We select a random position in the Sky, and the DM gamma-ray model for galaxy DDO 154 to estimate the expected limits on the relevant DM parameters. The fluctuations of the background counts follow a Poisson distribution. We repeat this process one thousand times and compute the average, and the $68\%$ and $95\%$ confidence intervals (CI) around the average value.

%Figures \ref{upperExpected} and \ref{lowerExpected} show the expected limits for five annihilation and decay channels, respectively. We also show, for reference, the exclusion limits obtained for DDO 154 galaxy and the combined analysis. Note that the DDO 154 galaxy shows a nearly $2\sigma$ overfluctuation in its observed background compared to expectations.

\section{Discussion and Results}\label{results}

Here we show the exclusion limits obtained for $31$ dIrr galaxies within the field of view of the HAWC Observatory. We present the limits for DM candidates with masses between $1~\text{TeV}$ and $100~\text{TeV}$. As explained in previous sections, the parameters of interest are the annihilation cross-section, $\langle\sigma_{\chi}v\rangle$, and the decay lifetime, $\tau_{\chi}$, of the dark matter particle. Here, we show and focus the discussion on the combined limits. The results for the individual limits are in Appendices \ref{apeAnna} and \ref{apeDec}.

\subsection{Annihilation}

\begin{figure}[tp]
\begin{center}
\includegraphics[width=0.75\linewidth]{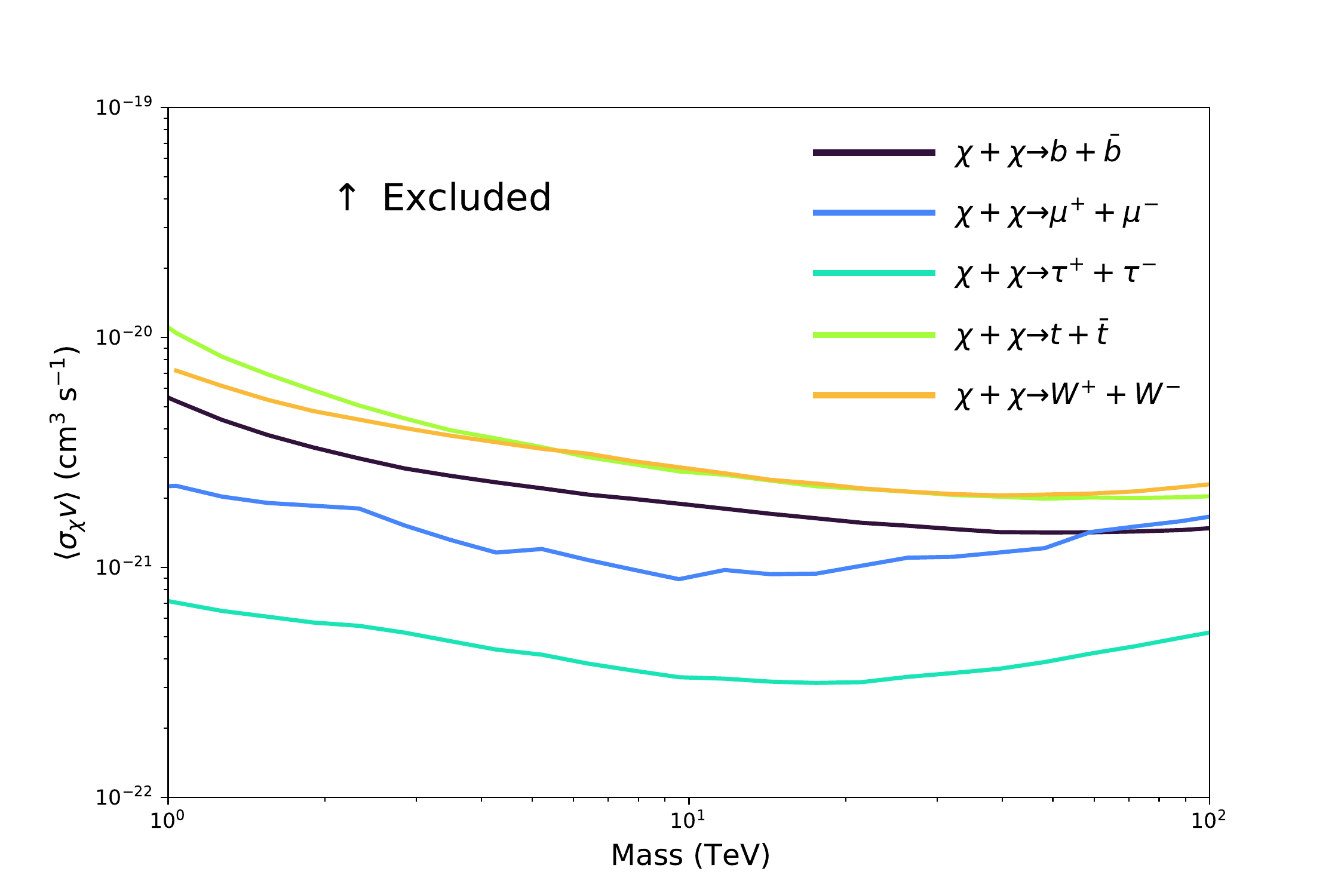}
\caption{The combined upper limits for all annihilation channels we considered in this study. The curves exclude the region above them. We computed the combined limit using the 31 dIrr galaxies in our sample.}
\label{upper_limits}
\end{center}
\end{figure}

The calculation of the combinned upper limits on the annihilation cross-section used all the galaxies in the sample. For simplicity, we assumed a point-source model for all galaxies. 
%Under this assumption, we have no problems with spatial correlations between the different regions of interest. 
We show results for five annihilation channels: $b,~\mu,~\tau,~t,~\text{and}~W$, see Figure \ref{upper_limits}. We observe that the most constraining limits are for the annihilation channel to $\tau$ lepton.

\begin{figure}[htp]
\centering
\includegraphics[width=0.765\textwidth]{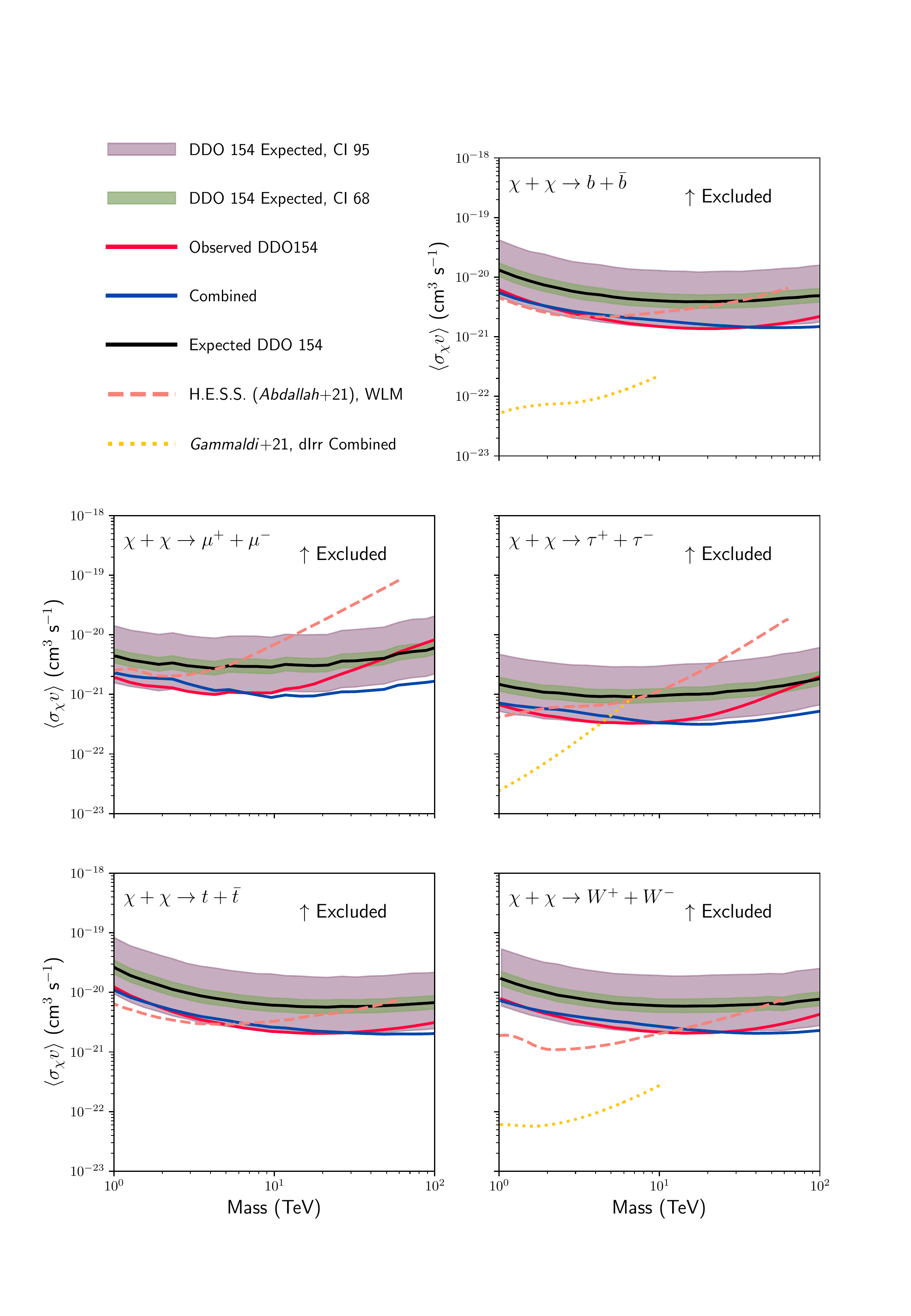}
\caption{Exclusion upper limits for DM annihilation cross-section for five channels. Top-Right: $b\bar{b}$ quarks, Middle-Left: $\mu^{+}\mu^{-}$ leptons, Middle-Right: $\tau^{+}\tau^{-}$ leptons, Bottom-Left: $t\bar{t}$ quarks Bottom-Right: $W^{+}W^{-}$ bosons. The blue and red lines show the upper limits obtained for the combined analysis and DDO 154 galaxy, respectively. The solid black line is the expected limit for the DDO 154 galaxy, and the color bands show $68\%$ and $95\%$ confidence intervals. The orange line shows the exclusion limits obtained from observations of the WLM galaxy with the H.E.S.S. array \citep{PhysRevD.103.102002}, and the yellow line shows the combined exclusion limits for 7 dIrr galaxies obtained with Fermi-LAT \citep{PhysRevD.104.083026}. The curves exclude the regions above them.}
\label{upperExpected}
\end{figure}

 We do not observe any difference between the combined limit (solid blue line) and the observed DDO 154 constraint (solid red line) in the lower mass range, see Figure \ref{upperExpected}. This effect is due to the large dispersion in the characteristics of the galaxies. In particular, galaxies with a small expected signal-to-background ratio contribute little to the combined limit.

The combined limit clearly shows a separation from the DDO 154 observed, though the magnitude of this effect depends on the channel. This separation starts at masses above $10~\text{TeV}$ for annihilation channels to $\mu$ and $\tau$ leptons; and above $30~\text{TeV}$ for $t$, $b$, and $W$ channels. Further studies are needed to fully address this effect, especially when including more galaxies in the combined analysis. However, the improvement for massive DM candidates shows the importance of combined analysis to obtain stronger constraining limits on DM parameters.

In addition, we calculate the expected limit for the DDO 154 galaxy. We obtain the average value, $68\%$, and $95\%$ confidence intervals after fluctuating the background events in the HAWC's maps, solid black line in Figure \ref{upperExpected}. We choose DDO 154 galaxy because it has the most constraining limit from all the galaxies in our sample and contributes most strongly to the combined limit.

We observe that, in general, the observed limit for galaxy DDO 154 (red line) is more restrictive than the expected DDO 154 limit (black line) for DM candidates with mass below $10~\text{TeV}$ ($30~\text{TeV}$) for $\mu$ and $\tau$ ($b,~t$, and $W$) channels. Fore more massive DM particles, the observed DDO 154 limit is closer to the expected DDO 154 exclusion limit. This behavior is related to the fact that the observed and expected counts of the background are approximately equal on the high $f_{\text{hit}}~\text{bins}$.

For lower masses, the discrepancy between the observed DDO 154 and the expected DDO 154 limits may be explained by a $2\sigma$ underfluctuation observed in the TS map for the region around galaxy DDO 154. This underfluctuation tell us that a DM model is likely less probable to describe the data observed in the galaxy DDO 154, and more consistent with the background-only hypothesis.

Finally, Figure \ref{upperExpected} shows also limits obtained from observation of other experiments. The data sets correspond to the exclusion limits obtained from observations of the galaxy WLM with H.E.S.S. (orange line), \citep{PhysRevD.103.102002}; and the combined limit for 7 dIrr galaxies within the Field of view of the Fermi-LAT experiment (yellow line), \citep{PhysRevD.104.083026}. We can observe that there is a nice complementarity between the three exclusion limits. Also, we observe that the HAWC combined limit is the most restrictive for dIrr galaxies for masses above $10~\text{TeV}$.

\subsection{Decay}

\begin{figure}[tp]
\centering
\includegraphics[width=0.75\linewidth]{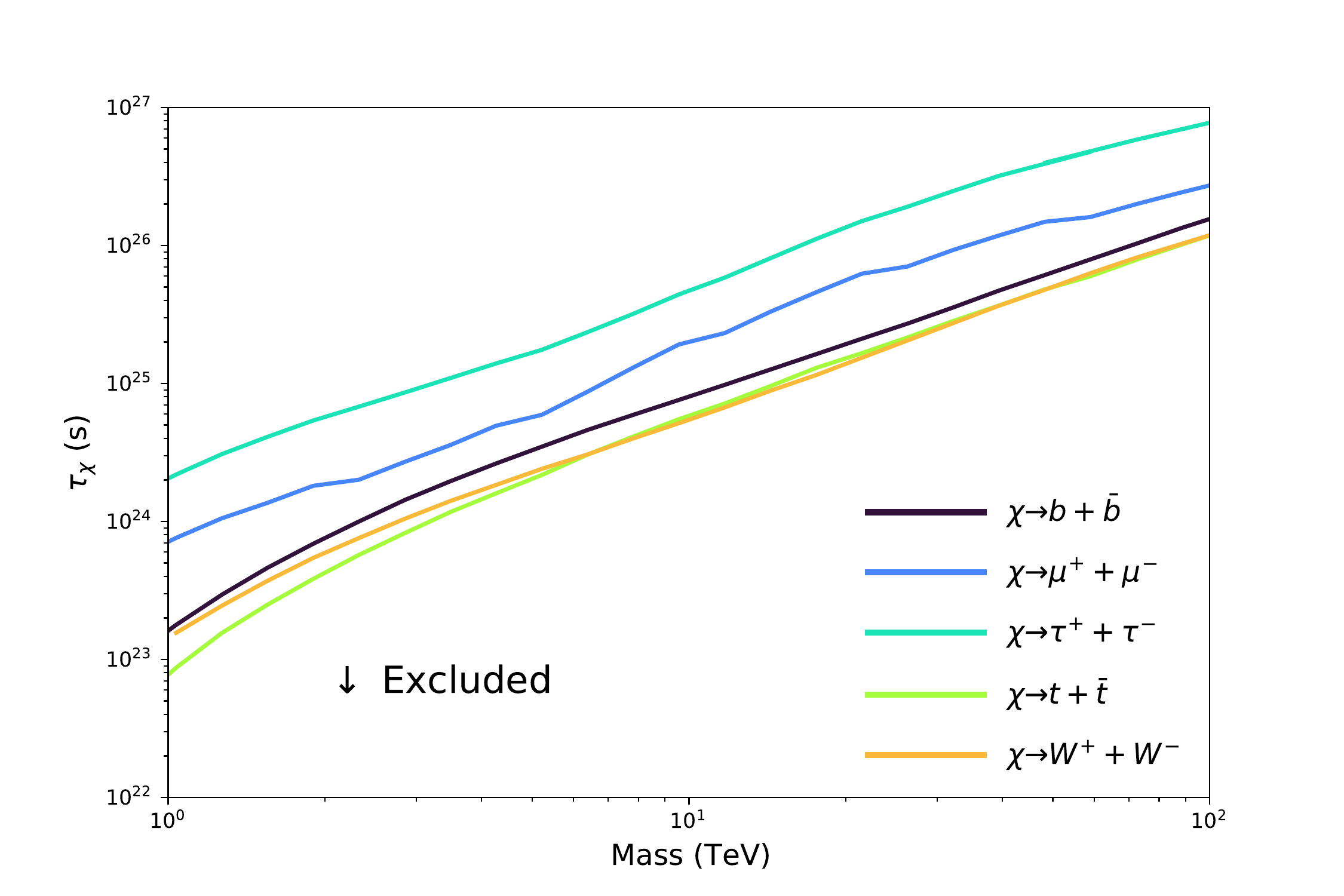}
\caption{The combined lower limits for all decay channels we considered in this study. The curves exclude the region below them. We computed the combined limit using the 31 dIrr galaxies in our sample.}
\label{lower_limits}
\end{figure}

\begin{figure}[htp]
\centering
\includegraphics[width=0.8\linewidth]{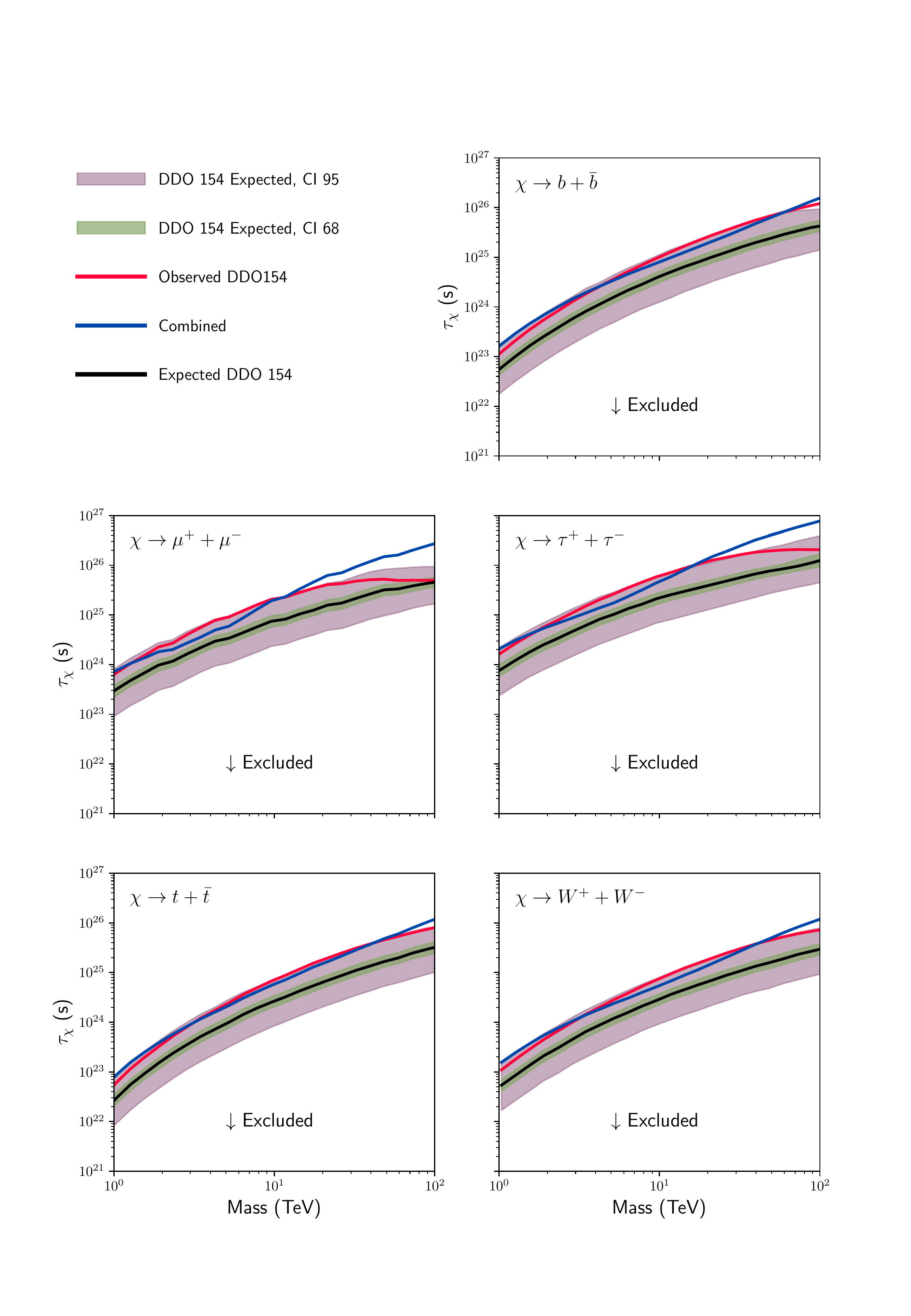}
\vspace{0.3cm}
\caption{Exclusion lower limits for DM decay lifetime for five channels. Top-Right: $b\bar{b}$ quarks, Middle-Left: $\mu^{+}\mu^{-}$ leptons, Middle-Right: $\tau^{+}\tau^{-}$ leptons, Bottom-Left: $t\bar{t}$ quarks Bottom-Right: $W^{+}W^{-}$ bosons. The blue and red lines show the lower limits obtained for the combined analysis and DDO 154 galaxy, respectively. The solid black line is the expected limit for the DDO 154 galaxy, and the color bands show $68\%$ and $95\%$ confidence intervals. The curves exclude the regions below them.}
\label{lowerExpected}
\end{figure}

While it is often supposed that WIMPs are stable particles, in some super-symmetric models, the DM candidate can decay to SM particles. Adding a term that breaks down R-parity allows this process. To match the DM relic density at the present epoch, the candidate must live much longer than the age of the Universe \cite{decayDM2015}. In super-symmetric models, the estimated lifetime is usually $\tau_{\chi}\thicksim10^{27}~\text{s}$ \citep{DecayingDMI,DecayingDMII}. Figure \ref{lower_limits} shows the lower combined limits obtained for our galaxy sample, and Figure \ref{lowerExpected} also shows the expected limits for galaxy DDO 154. As in the annihilation case, we observe that the most restrictive combined limit is for the decay channel to $\tau$ leptons, where we obtain a lifetime $>10^{26}~\text{s}$ for masses above $\thicksim20~\text{TeV}$. We note the same behavior for the combined and DDO 154 limits, with an improvement in the high masses regime showing the importance of combined analysis for these targets.

\subsection{Comparison with dwarf spheroidal galaxies}
\label{comparison}

We have obtained the exclusion limits under the assumption that dIrr galaxies do not show evidence of astrophysical processes contributing to the gamma-ray flux. The same hypothesis is valid for dSph galaxies, and we can compare the results between both target classes. For the comparison, we used the dSph limits reported in \cite{hawcdsph}. There exists a classification of the dSph galaxy population based on the number of star members in the galaxy: classical and ultra-faint dSph galaxies. The designation also refers to the quality of available kinematical data, which impacts the level of uncertainty on the astrophysical factors: better data, smaller uncertainty. We used the DDO 154 and combined limits from the dIrr sample to observe possible similarities between the populations. The objective is to encourage further studies combining these galaxy populations and put more constraining limits in the DM parameter space. 

\begin{figure}[htp]
\centering
\includegraphics[width=0.85\linewidth]{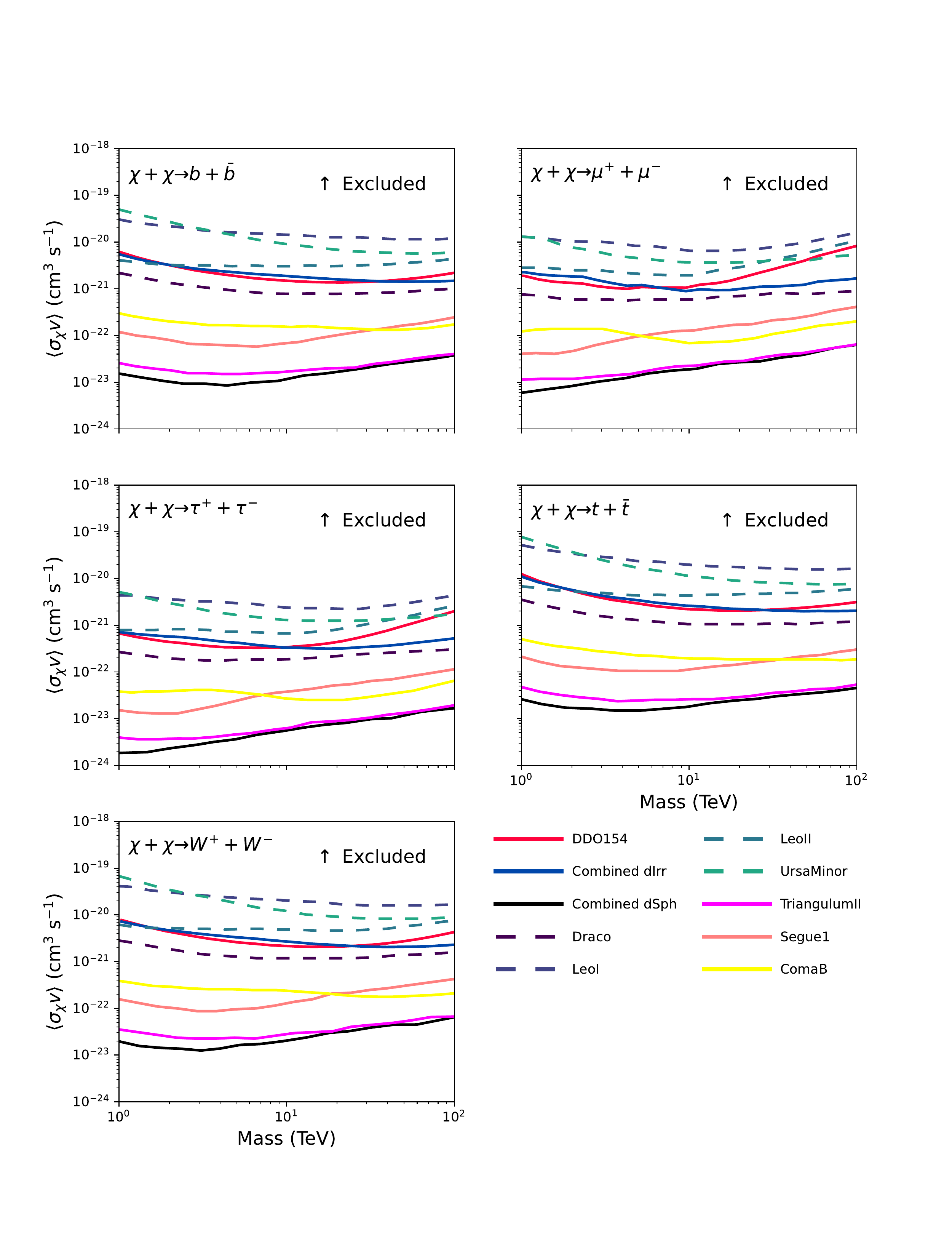}
\vspace{0.3cm}
\caption{Exclusion upper limits for DM annihilation cross-section for dIrr and dSph Galaxies within the field of view of the HAWC Observatory for five annihilation channels: Top-Left: $b\bar{b}$ quarks, Top-Right: $\mu^{+}\mu^{-}$ leptons, Middle-Left: $\tau^{+}\tau^{-}$ leptons, Middle-Right: $t\bar{t}$ quarks, Bottom-Left: $W^{+}W^{-}$ bosons. The black and red lines show the upper limits obtained for the dIrr combined analysis and DDO 154 galaxy, respectively. The exclusion limits for ultrafaint (solid lines) and classical (dashed lines) dSph galaxies are shown. The region above the curves is excluded.}
\label{annihilationcomparison}
\end{figure}

\begin{figure}[htp]
\centering
\includegraphics[width=0.85\linewidth]{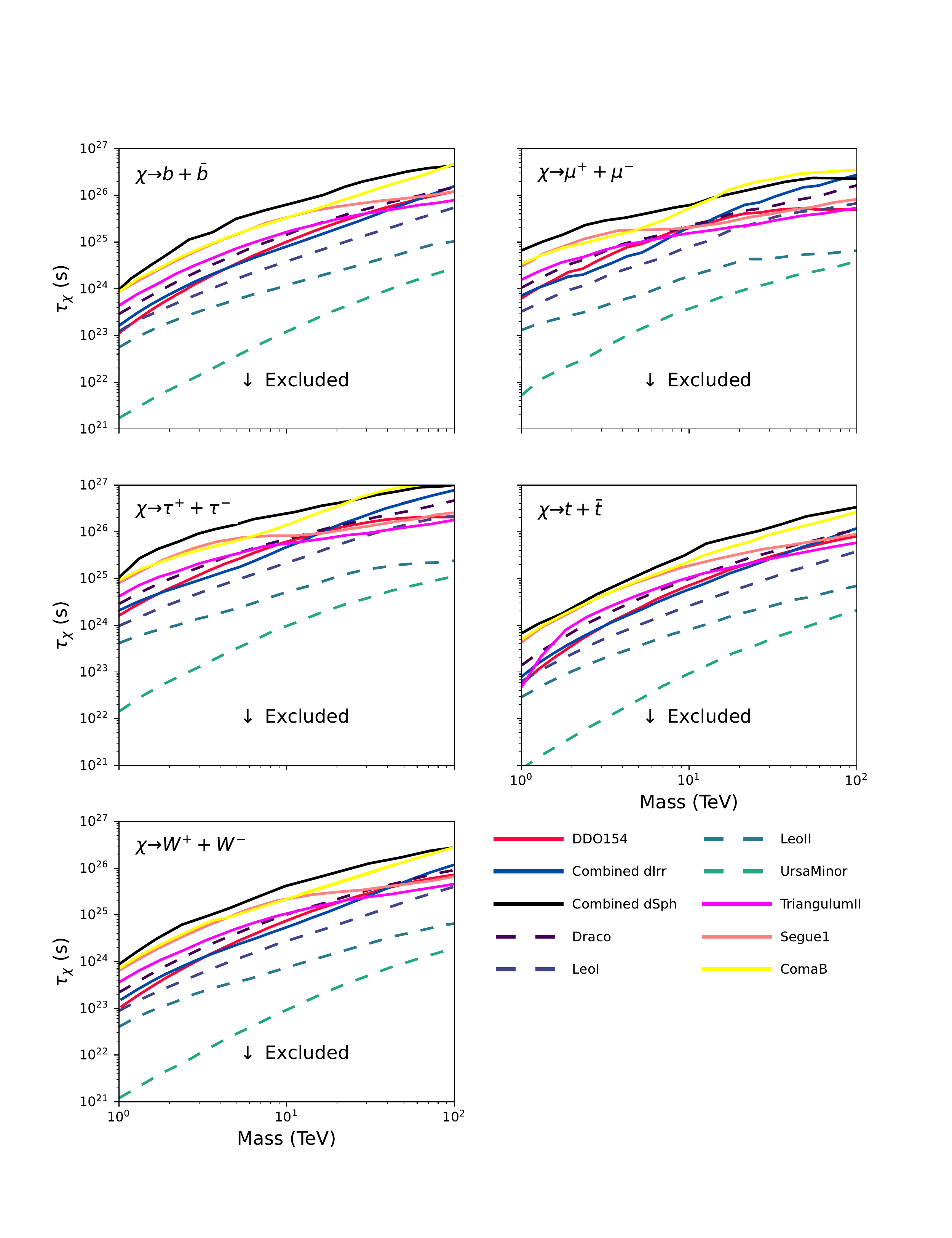}
\vspace{0.3cm}
\caption{Exclusion lower limits for DM decay lifetime for dIrr and dSph Galaxies within the field of view of the HAWC Observatory for five decay channels: Top-Left: $b\bar{b}$ quarks, Top-Right: $\mu^{+}\mu^{-}$ leptons, Middle-Left: $\tau^{+}\tau^{-}$ leptons, Middle-Right: $t\bar{t}$ quarks, Bottom-Left: $W^{+}W^{-}$ bosons. The black and red lines show the upper limits obtained for the dIrr combined analysis and  DDO 154 galaxy, respectively. The exclusion limits for ultrafaint (solid lines) and classical (dashed lines) dSph galaxies are shown. The region below the curves is excluded.}
\label{decaycomparison}
\end{figure}

Figures \ref{annihilationcomparison} and \ref{decaycomparison} show the comparison between dIrr and dSph galaxies for dark matter particles annihilating or decaying to five different channels. The black and red lines represent the dIrr combined and DDO 154 limits, respectively. The dotted lines show the exclusion limits for classical (Draco, Leo I, Leo II, and Ursa Minor) dSphs and the solid lines for ultrafaint (Triangulum II, Segue I, and Coma Berenices) dSph galaxies. We observe that, in general, the limits from dIrr galaxies are less restrictive than the ultrafaint dSphs up to two orders of magnitude for annihilation and one order of magnitude for decay. The reason is that ultrafaint galaxies have astrophysical factors larger than our sample. Also, one must remember that the associated uncertainty is considerably higher for ultrafaint galaxies. For example, in recent studies \citep{scalingrelations2018}, the astrophysical factor obtained for Triangulum II is an upper limit, then the actual constraint could be less restrictive. We also observe that the combined limit for dIrr galaxies is comparable (has the same order of magnitude) to limits from classical dSph galaxies. The reason is that the astrophysical factors between the two populations are very similar. Note that this does not imply that both galaxy populations have or share the same properties (dIrr galaxies are farther than spheroidals, but dIrr are more massive, for example).

We also observe that the combined limits of dIrr and dSph galaxies have similar behavior. The improvement is relatively small in both populations, probably caused by background fluctuations in galaxies in declinations bands where HAWC has lower sensitivity; or by an effect due to the extension of the target. They both impact the calculation and minimization of the likelihood function. As explained in section \ref{JLana}, the first scenario can lead to a decrease for the best limit when different galaxies in the sample have background counts differing from each other. In our case, one possible solution could be only to use galaxies within declination bands around the target that contributes the most to the combined limit (DDO 154). Also, including only targets with similar expected signal intensity (or J- and D- factors) may help to increase the combined limit. We will explore this possibility in next analyses.

\section{Conclusions}\label{conclusions}

We have shown that dIrr galaxies are a new suitable population of objects to perform indirect DM searches by extended-array experiments, like HAWC Observatory. This is possible because of the environmental conditions of these galaxies, with low SFRs and low population of massive stars. We computed the exclusion limits at 95\% C.L. for annihilation and decay of generic WIMPs with masses between 1 and 100 TeV. In both cases, the best limits are obtained for the galaxy DDO154, located in a declination band where HAWC has good sensitivity. The computed limits were obtained under the assumption that the DM density is described by a Burkert profile and not enhancement by substructure was considered, but it is possible that the presence of a central black hole in these galaxies can modify the slope of the DM distribution in the central region of dIrrs. This would lead to an enhancement in the annihilation exclusion limit, at least for one or two orders of magnitude, according to the dynamic history of the host galaxy \citep{BHdSph}, and the fact that the profile is modified via the rotational motion of the galaxy \citep{oman2018}, leading to more cusped profiles, like the NFW profile \citep{nfw1}. More studies in this approach are planned for an upcoming publication. Also, we observe more dIrr galaxies in the Local Universe ($\thicksim11~\text{Mpc}$), and we can obtain their mass distribution from rotation curve data. Therefore, combined analysis as described in this work will lead to more stringent constraints of DM candidates. Moreover, given the established universality of their mass distribution, useful constraints for the DM cross sections can be obtained even for a limited amount of observations for each of them. We also show that our exclusion limits are comparable to the limits obtained for classical dSph galaxies, so future studies may consider a possible joint analysis between the two populations to be able to constrain even more the properties of DM candidates. 

%%%%%%%%%%%%%%%%%%%%%%%%%%%%%%%%%%%%%

The goal of the paper is to show the capabilities of the HAWC Observatory to use large populations of galaxies to constraints parameters to DM models through the search of gamma-ray signatures of annihilation or decay of DM particles. Here we used a sample of dIrr galaxies. However, the present analysis is the starting point for more detailed studies. In this analysis we assumed that the DM induced gamma-ray emission is described by a point source, but this is not the case for galaxies like IC 10, IC 1613 and NGC 6822, with extension across the sky of $\gtrsim2$ degrees (see Table \ref{dIrrSample}). Using an extended emission model should lead to more realistic constraints. Furthermore, with an analysis comprising a larger period of time and improved HAWC energy estimators, we can improve the constraints, both in annihilation and decay. We may also obtain limits to DM candidates masses in the high-end range of masses expected for WIMPs \citep{ENQVIST1991367,smirnov_2019}. Therefore, we will be probing the hypothesis about the thermal production of WIMPs in the early Universe. Note that this can be an indication of other interesting escenarios as multicomponent or light DM.

Secondly, as we pointed at the beginning of this section, dIrr galaxies are abundant in the local Universe and combined analysis using more targets should be adressed. Moreover, as we discussed in Section \ref{comparison}, this also give us the opportunity to test different strategies of combining the data from dIrr galaxies. For example, in order to reduce the background fluctuations we may use targets with similar expected signal-to-background ratio in the same declination band to check if the combined limit can be improved.

Finally, combined limits using not only the dIrr galaxies population, but also classical dSph galaxies should help to increase the contraints of the different dark matter parameters.

\section*{Acknowledgments}
We acknowledge the support from: the US National Science Foundation (NSF); the US Department of Energy Office of High-Energy Physics; the Laboratory Directed Research and Development (LDRD) program of Los Alamos National Laboratory; Consejo Nacional de Ciencia y Tecnolog\'ia (CONACyT), M\'exico, grants 271051, 232656, 260378, 179588, 254964, 258865, 243290, 132197, A1-S-46288, A1-S-22784, c\'atedras 873, 1563, 341, 323, Red HAWC, M\'exico; DGAPA-UNAM grants IG101320, IN111716-3, IN111419, IA102019, IN110621, IN110521, IN102223; VIEP-BUAP; PIFI 2012, 2013, PROFOCIE 2014, 2015; the University of Wisconsin Alumni Research Foundation; the Institute of Geophysics, Planetary Physics, and Signatures at Los Alamos National Laboratory; Polish Science Centre grant, DEC-2017/27/B/ST9/02272; Coordinaci\'on de la Investigaci\'on Cient\'ifica de la Universidad Michoacana; Royal Society - Newton Advanced Fellowship 180385; Generalitat Valenciana, grant CIDEGENT/2018/034; The Program Management Unit for Human Resources \& Institutional Development, Research and Innovation, NXPO (grant number B16F630069); Coordinaci\'on General Acad\'emica e Innovaci\'on (CGAI-UdeG), PRODEP-SEP UDG-CA-499; Institute of Cosmic Ray Research (ICRR), University of Tokyo, H.F. acknowledges support by NASA under award number 80GSFC21M0002. We also acknowledge the significant contributions over many years of Stefan Westerhoff, Gaurang Yodh and Arnulfo Zepeda Dominguez, all deceased members of the HAWC collaboration. Thanks to Scott Delay, Luciano D\'iaz and Eduardo Murrieta for technical support. Viviana Gammaldi contribution to this work has been supported by \textit{Juan de la Cierva-Incorporaci\'on} IJC2019-040315-I grants, by the grants PGC2018-095161-B-I00, CEX2020-001007-S, PID2021-125331NB-I00 all funded by MCIN/AEI/10.13039/501100011033 and by ``ERDF A way of making Europe’'.%work has been supported by JUAN DE LA CIERVA-FORMACI\'ON FJCI-2016-29213, by the Spanish Agencia Estatal de Investigaci\'on through the grants FPA2015-65929-P (MINECO/FEDER, UE) and IFT Centro de Excelencia Severo Ochoa SEV-2016-0597, by INFN project QGSKY, by the Agencia Estatal de Investigaci\'on (AEI) y al Fondo Europeo de Desarrollo Regional (FEDER) FIS2016-78859-P(AEI/FEDER, UE) and partially by the H2020 CSA Twinning project No.692194 ORBI-T-WINNINGO. VG also acknowledges the support of the Spanish Red Consolider MultiDark FPA2017-90566-REDC.
% \end{acknowledgments}


\bibliography{sample631}{}
\bibliographystyle{aasjournal}



%%%%%%%%%%%%%%%%%%%%%%%%%%%%%%%%%%%%%%%%%%

\appendix
\section{Individual Exclusion Limits for Annihilation}\label{apeAnna}

    Here we present the individual exclusion limits for all the targets in our sample. For completeness, we also show the combined limit as described in the Section \ref{JLana}.
    
\begin{figure}[t!]
    % \vspace{-0.2cm}
    % \captionsetup{width=0.7\linewidth}
    \centering
    \includegraphics[width=0.8\linewidth]{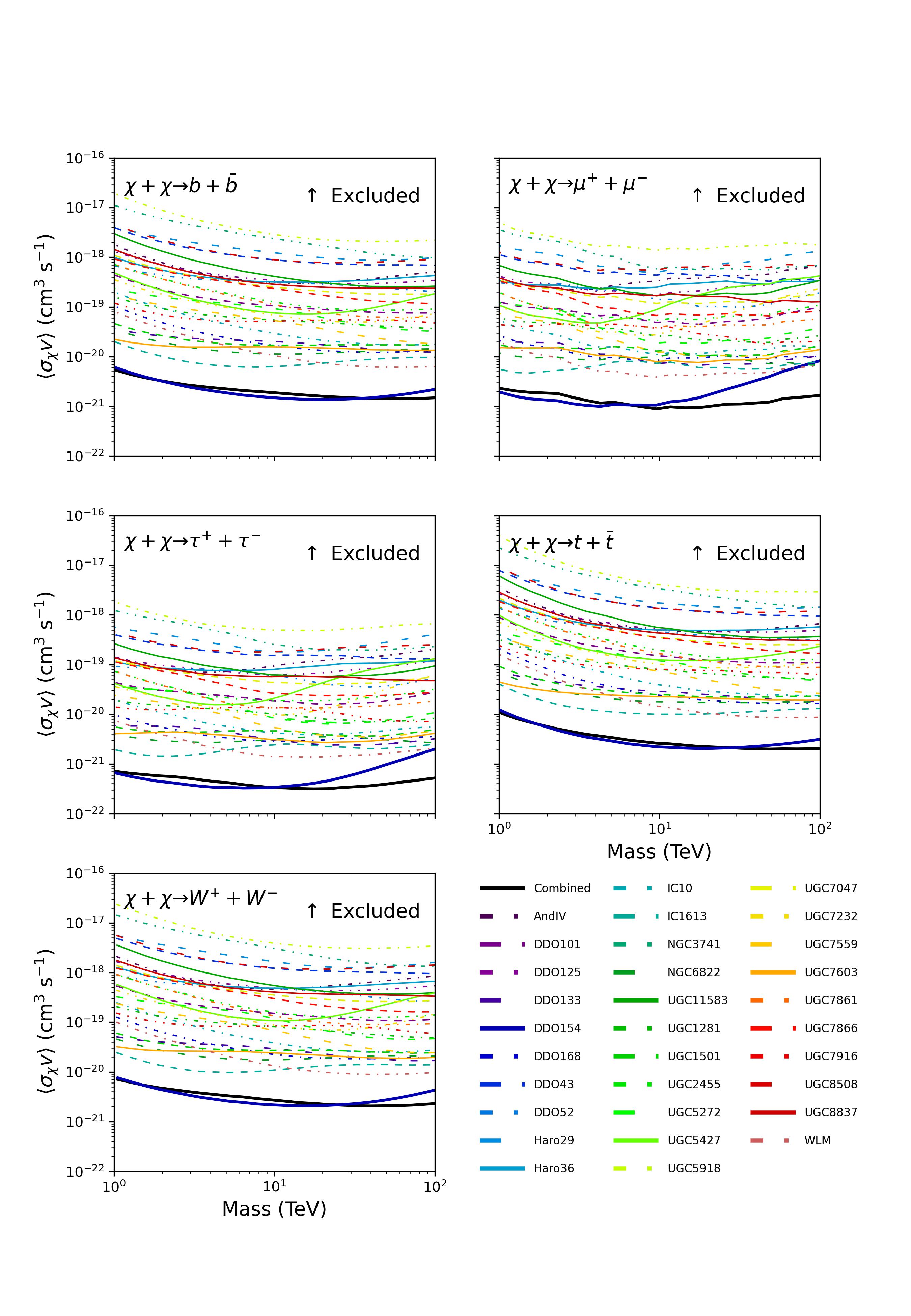}
    % \vspace{0.3cm}
    \caption{Exclusion upper limits for DM annihilation cross section for dIrr Galaxies in the field of view of the HAWC Observatory for five annihilation channels. The solid-black line in the figures for every annihilation channel shows the combined analysis using all the galaxies in the sample. The region above the curves is excluded.}
    \label{AnnaInd}
    \end{figure}

    In this section we show the exclusion limits for the 31 dIrr galaxies in our sample, see Figure \ref{AnnaInd}. We observe that the best individual constraining limit is for the galaxy DDO154. While its astrophysical factor is not the largest in our sample, DDO154 is located in a declination band where HAWC has good sensitivity. For completeness, we also show the combined limit in Figure \ref{AnnaInd}. We can also observe that the DDO154 and combined limits are similar for energies below $10~\text{TeV}$. For channels such as $\mu$ and $\tau$, the combined limit shows an improvement above $10~\text{TeV}$. For the other annihilation channels, the improvement in the combined limit only appears at the high-energy regime. An explanation could be the high background rejection at high energies by HAWC, providing cleaner sample data for all the dIrr galaxies.
    
    As we pointed out in Section \ref{results}, another explanation for the null difference between the combined and the DDO154 limits is the large scatter in properties of all galaxies in the sample. In particular, the individual backgrounds should contribute to this effect.

\section{Individual Exclusion Limits for decay}\label{apeDec}

\begin{figure}[t!]
    % \vspace{-0.5cm}
    % \captionsetup{width=0.7\linewidth}
    \centering
    \includegraphics[width=0.8\linewidth]{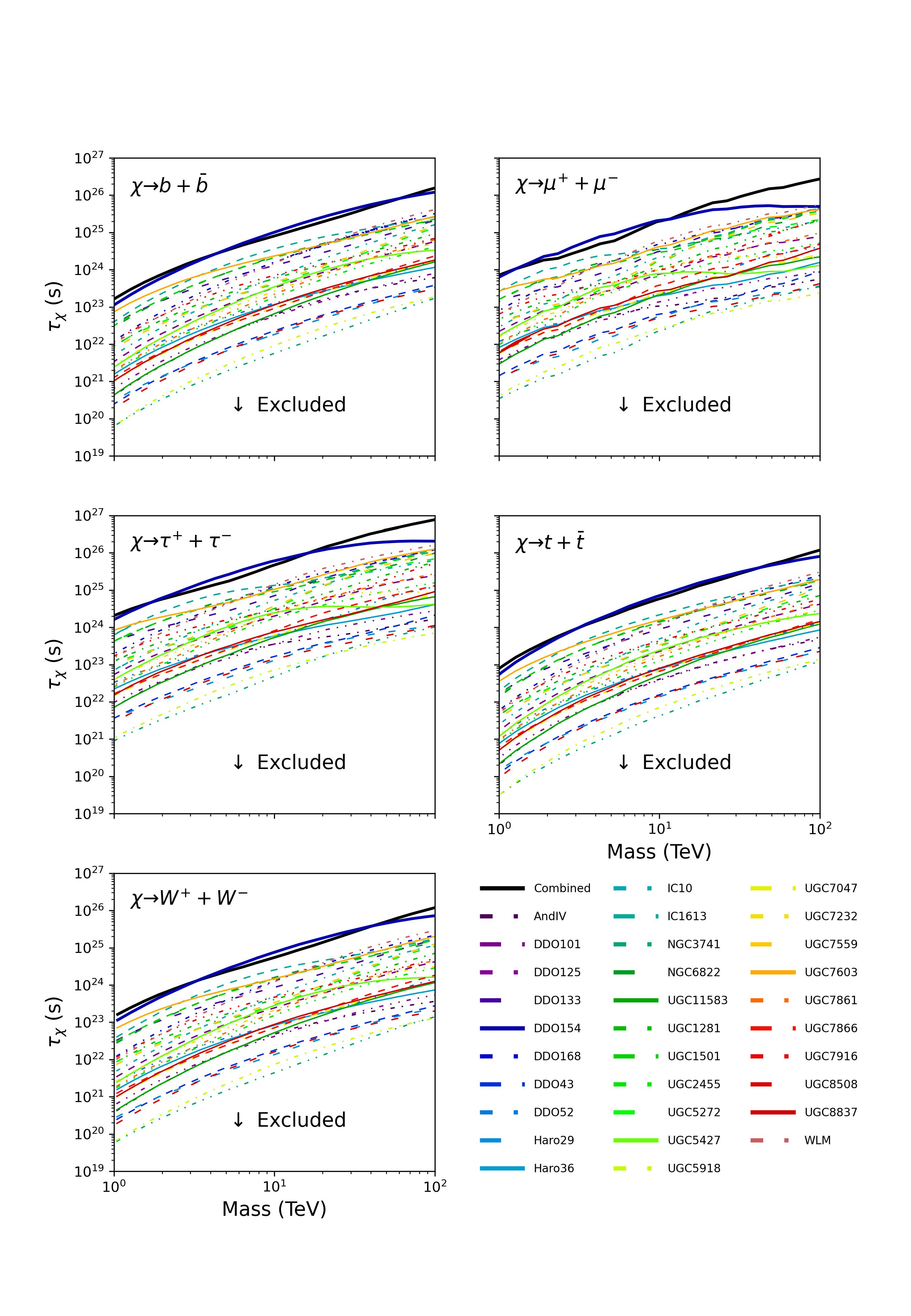}
    % \vspace{0.3cm}
    \caption{Exclusion lower limits for DM decay lifetime for dIrr Galaxies in the field of view of the HAWC Observatory for five annihilation channels. The solid-black line in the figures for every decay channel shows the combined analysis using all the galaxies in the sample. The region below the curves is excluded.}
    \label{DecInd}
    \end{figure}

    In this section we show the exclusion limits for the 31 dIrr galaxies in our sample (see Figure \ref{DecInd}). Again, we observe that the best individual limit is for the galaxy DDO 154. As in the annihilation case, the probable reason is DDO 154 is in a favorable declination band. For completeness, we also show the combined limit. We observe the same features between the combined and the DDO 154 limits. As we discuss in Section \ref{results}, the decay limits (individual and combined) for dIrr galaxies are competitive with those obtained for other targets as dSph galaxies. The main reason is the high mass of dIrr galaxies. 

\end{document}